\documentclass[%
 reprint,
superscriptaddress,
 amsmath,amssymb,
aps,
]{revtex4-2}

\usepackage{graphicx}
\usepackage{dcolumn}
\usepackage{bm}

\usepackage{amsmath}
\usepackage{braket}

\usepackage{calligra}
\DeclareMathAlphabet{\mathcalligra}{T1}{calligra}{m}{n}
\DeclareFontShape{T1}{calligra}{m}{n}{<->s*[2.2]callig15}{}

\usepackage{color}

\begin{document}

\preprint{APS/123-QED}

\title{A locally \emph{ab initio} computational framework for arbitrary incommensurate materials interfaces}

\author{Drake Niedzielski}
\email{dan98@cornell.edu}
\affiliation{%
 Department of Physics, Cornell University, Ithaca, NY 14853, USA.
}
\author{Tomás A.~Arias}
\email{taa2@cornell.edu}
\affiliation{%
 Department of Physics, Cornell University, Ithaca, NY 14853, USA.
}

\date{\today}

\begin{abstract}

Incommensurate materials interfaces constitute a broad and technologically important class of systems, yet their lack of shared periodicity limits predictive and computationally efficient first-principles electronic-structure methods. Here we introduce a scalable computational framework for constructing locally \emph{ab initio} electronic Hamiltonians for arbitrary materials interfaces. Our approach exploits the nearsightedness of Wannier Hamiltonian matrix elements, enabling their systematic extrapolation and interpolation across interlayer registries. This strategy yields transferable Hamiltonians that retain first-principles accuracy while bypassing the need for prohibitively large commensurate supercells or Moir\'e approximations. We validate the framework on quasicrystalline $30^\circ$ twisted bilayer graphene, reproducing experimentally observed spectral features including mirrored Dirac cones and minigaps at avoided crossings arising from generalized interlayer scattering. We further predict quasiperiodic flat-band states in experimentally accessible doping regimes. By enabling predictive electronic-structure calculations across structurally incommensurate interfaces, this framework establishes a practical route to first-principles exploration of emergent interfacial phenomena.

\end{abstract}

\maketitle 

\section{Introduction}

The discovery of strongly correlated phenomena in twisted two-dimensional materials, most famously the emergence of flat bands and superconductivity in magic-angle twisted bilayer graphene~\cite{cao_unconventional_2018}, has established interlayer twist as a powerful means of engineering electronic structure and emergent quantum states \cite{andrei_marvels_2021, mak_semiconductor_2022}. More broadly, twisted and lattice-mismatched heterostructures form a rapidly expanding class of incommensurate materials interfaces whose electronic properties are often unexpectedly different from that of their constituent layers, necessitating accurate theoretical description. Theoretical treatment of such systems is fundamentally challenging because relative twist or lattice mismatch between layers breaks global crystal symmetry, precluding a common supercell description required by conventional \emph{ab initio} methods. Though not \emph{ab initio}, continuum models developed by MacDonald, Koshino, and others successfully capture low-energy physics in the small-angle, weak-coupling regime, where long-wavelength Moir\'e periodicity enables effective descriptions of interlayer coupling\cite{bistritzer_moire_2011, koshino_effective_2020}. However, many experimentally relevant interfaces lie beyond this limit, where strong coupling or large twist angles invalidate the Moir\'e approximation, leaving predictive first-principles theory an outstanding challenge.

Incommensurate interfaces span a wide range of bonding environments and lattice symmetries, including strongly coupled misfit rare-earth rocksalt transition-metal dichalcogenide systems~\cite{ng_misfit_2022}, twisted oxides\cite{pryds_twisted_2024}, and high-angle twisted quasicrystals. Twisted cuprate heterostructures have emerged as a platform of growing interest, with twist angles approaching $45^\circ$ proposed as a route for engineering unconventional superconductivity through interfacial interaction \cite{zhao_time-reversal_2023}. In such systems, strong interlayer bonding and large twist angles, the latter of which drives the Moir\'e periodicity to length scales comparable to the constituent lattice constants, invalidate both the weak-interaction and Moir\'e approximations, respectively. Quasicrystals are the extreme manifestation of this high twist-angle regime. Notably, $30^\circ$ twisted bilayer graphene ($30^\circ$-TBG) forms a dodecagonal quasicrystal, exhibiting long-range quasiperiodicity without translational order, and hosts subtle ``mirrored" electronic states \cite{yao_quasicrystalline_2018, park_higher-order_2019, moon_quasicrystalline_2019, pezzini_30-twisted_2020, hamer_moire_2022}. This system has been proposed to host quasicrystalline superconductivity, with enhanced critical temperature, and flat-band formation at dopings far from the Fermi level driven by interlayer hybridization and real-space localization~\cite{ghadimi_quasiperiodic_2025}. A predictive \emph{ab initio} framework is therefore needed to study these subtle and important electronic phenomena in incommensurate layered materials.

$30^\circ$-TBG, in particular, provides a stringent benchmark for first-principles theories of incommensurate interfaces through reproduction of its experimentally observed mirrored Dirac cones. These features are thought to arise from generalized interlayer Umklapp scattering between the reciprocal lattices of the two graphene layers, producing replica Dirac dispersions at momenta not related by crystalline symmetry. Experimentally resolved through angle-resolved photoemission spectroscopy (ARPES), these subtle mirrored features appear much fainter than the main Dirac cones \cite{ahn_dirac_2018, yao_quasicrystalline_2018, hamer_moire_2022}. Capturing these signatures thus presents a substantial theoretical challenge. Continuum and tight-binding approaches can reproduce mirrored cones when generalized scattering processes are explicitly assumed, but in such treatments the features are effectively imposed through model construction rather than emerging directly from a first-principles description of the incommensurate interface \cite{moon_quasicrystalline_2019}. Moreover, as seen in ARPES, intersections between mirrored and primary cones produce hybridization-induced avoided crossings that generate minigaps, whose existence depends on interlayer coupling \cite{yao_quasicrystalline_2018, hamer_moire_2022}. Beyond these dispersive features, quasiperiodic interlayer hybridization in these systems has been associated with singularities in the density of states and proposed flat-band formation at energies far from the Fermi level \cite{moon_quasicrystalline_2019, hamer_moire_2022}. The simultaneous reproduction of mirrored cones and  their hybridization induced minigaps therefore provides a stringent test of predictive frameworks for incommensurate electronic structure.

A substantial body of theoretical work has sought to address the electronic structure of incommensurate heterostructures, yet predictive modeling remains a central challenge. In the absence of a common supercell, one must typically construct large, artificially strained commensurate approximants to preserve periodic boundary conditions across both layers. Minimizing artificial strain, however, often requires supercells so large that conventional \emph{ab initio} calculations, implemented using density functional theory (DFT) \cite{kohn_self-consistent_1965}, become computationally prohibitive. Continuum and empirical tight-binding approaches offer efficient alternatives, but rely on weak interlayer coupling, assume the validity of the Moir\'e approximation, and require new model parameterizations for each distinct material combination. \emph{Ab initio} cluster--supercell extrapolation techniques~\cite{niedzielski_unmasking_2025,gerber_ab_2020} maintain first-principles accuracy while mitigating artificial strain by treating one material as a finite cluster and the other as a periodic substrate. While effective for observables expressed as energy-integrated spectral functions, finite cluster size introduces quantization effects that limit the resolution of sharp momentum, real-space delocalized, features ---including gaps arising from avoided interlayer band crossings---particularly when both layers contribute comparably in the relevant energy range, as in 30$^\circ$-TBG. A predictive framework for arbitrary incommensurate interfaces must therefore treat all layers on equal footing, without invoking Moir\'e periodicity, while remaining both computationally tractable and transferable across material systems.

In this work, we introduce a general, fully \emph{ab initio} computational framework for incommensurate materials interfaces based on locally constructed Wannier Hamiltonians. Building on our previous Mismatched INterface Theory (MINT)~\cite{niedzielski_unmasking_2025,gerber_ab_2020}, an \emph{ab initio} cluster–supercell extrapolation technique, we demonstrate that Wannier Hamiltonian matrix elements exhibit nearsightedness and a smooth dependence on local atomic environment, enabling us to reconstruct the electronic structure of incommensurate interfaces from only a handful of convergent local environment calculations. To avoid explicit first-principles calculations for every local atomic configuration, we develop interpolation schemes for both intra- and interlayer hoppings as smooth, registry-dependent functions defined over the unit cell, obtained through local regression. Our approach simultaneously interpolates Hamiltonian matrix elements across local environments, suppresses numerical noise, and systematically extrapolates toward the bulk limit through inverse-size scaling. Applied to quasicrystalline $30^\circ$-TBG, our framework reproduces the experimentally observed mirrored Dirac cones, higher-order scattered replicas, and hybridization-induced minigaps at interlayer band crossings. It further predicts spectral features associated with quasiperiodic interlayer hybridization, including flat-bands in the electronic structure, and allows for the efficient study of electronic wavefunction localization via the inverse participation ratio (IPR). By enabling the local construction of predictive, first-principles Hamiltonians, without recourse to commensurate supercells or the Moir\'e approximation, our framework establishes a general route to first-principles electronic-structure modeling across arbitrary incommensurate interfaces.

\section{Results}

\subsection{Local Registry Description of Incommensurate Interfaces}

\begin{figure}
\includegraphics[width=\linewidth]{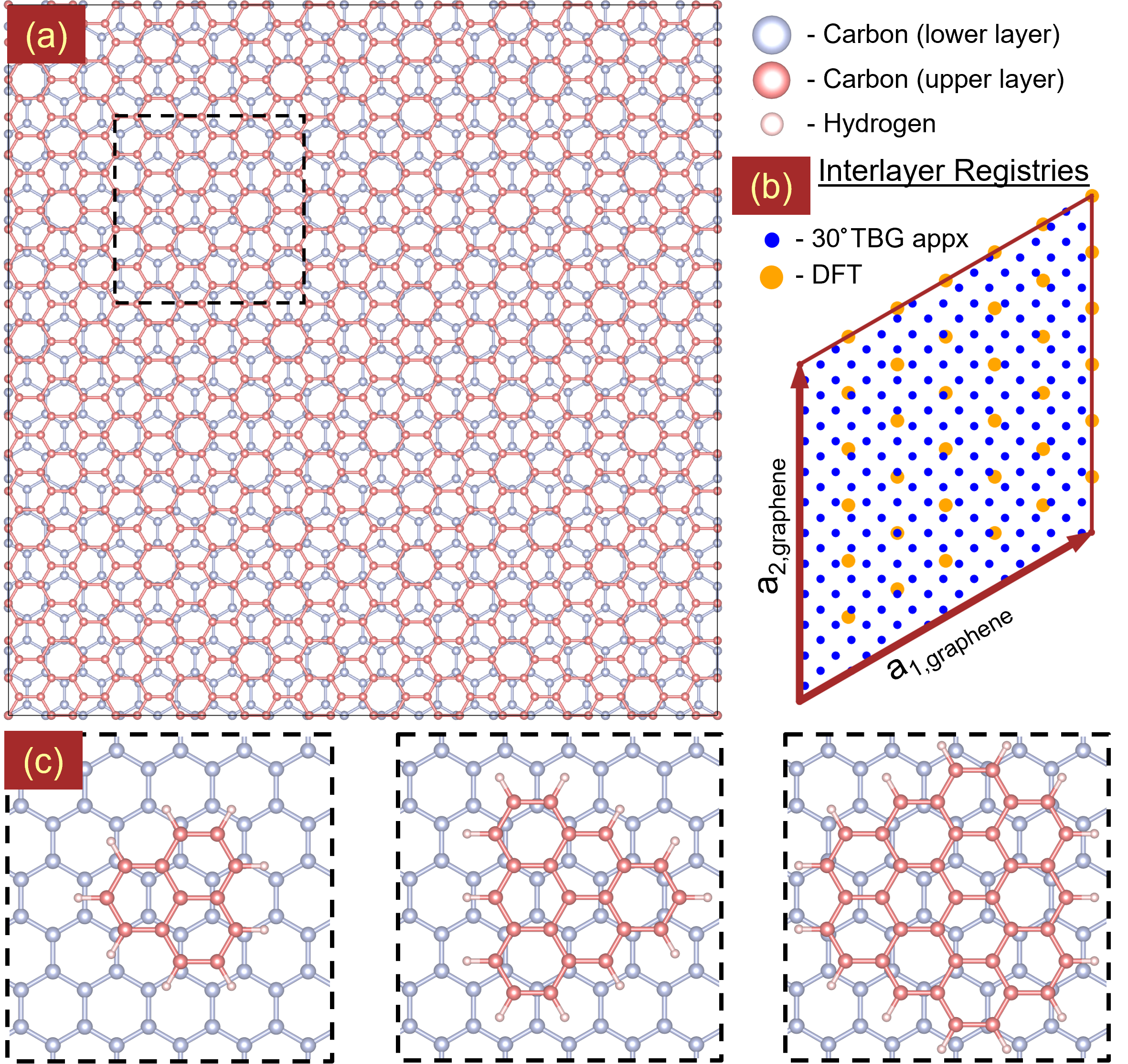}
\caption{Atomic-scale sampling of local interlayer registries in $30^{\circ}$ twisted bilayer graphene. (a) Commensurate approximant of $30^{\circ}$-TBG containing 1672 atoms. (b) Local interlayer registries of all atomic environments in this 1672 atom system (blue dots) and a sparser $6\times6$ sampling (orange dots) at which we perform DFT calculations. (c) Three elements of MINT-proxy sequence whose structures converge to the local atomic environment of the bulk system (black dashed-line box in (a)). Atomic visualizations performed with VESTA \cite{momma_vesta_2011}.}
\label{fig:MatEx}
\end{figure}

Incommensurate bilayers are commonly approximated by large commensurate supercells obtained through continued-fraction approximations to the lattice constant ratio. While this requires the introduction of artificial strain, when made sufficiently large, such approximants can minimize this artificial strain while preserving correct boundary conditions on both layers. Fig.~\ref{fig:MatEx}(a) shows the fourth commensurate approximant of $30^\circ$-TBG containing 1672 atoms which results in only $\sim0.3\%$ artificial strain. This finite approximation, however, already lies beyond the practical limits of density functional theory, whose computation time scales as the cube of the system size. Solving this fundamental issue requires a different structural representation: one that does not depend on approximate commensuration but instead divides the incommensurate system into smaller, more manageable, pieces. We therefore adopt a local environment description of the incommensurate structure based on the relative stacking of the two layers.

The concept of a local environment is well established in the field of aperiodic solids, including quasicrystals, \cite{massatt_electronic_2017, ghadimi_mean-field_2020, ghadimi_quasiperiodic_2025, liu_learning_2025} as it remains well-defined even in the absence of crystal symmetry. In our approach, the environment of an atom at position $\vec{r}_A$ belonging to layer A, is described by a \emph{local interlayer registry}, $\vec{\mathcalligra r}(\vec{r}_A)$. Here, $\vec{\mathcalligra r}(\vec{r}_A)$ is defined as the displacement from the atom at $\vec{r}_A$ to the closest lattice vector of layer B. Note that the interlayer registry $\vec{\mathcalligra r}$ is always defined within the unit cell of lattice B. By calculating the registry with respect to the nearest unit cell of layer B rather than a distant origin, this definition is robust to atomic strain (see Methods). In the absence of strain, our definition becomes equivalent to similar local environment descriptors employed by references \cite{ghadimi_quasiperiodic_2025} and \cite{massatt_electronic_2017}. While a truly incommensurate interface samples the full continuum of registries, a finite approximant samples only a discrete subset. This is illustrated in Fig.~\ref{fig:MatEx}(b), whose blue dots show the discrete set of interlayer registries present in the fourth commensurate approximant Fig.~\ref{fig:MatEx}(a). The convergence of material properties with increasing commensurate supercell size then becomes analogous to the convergence of discrete Brillouin zone sampling in conventional DFT.

\subsection{Local Electronic Structure}

Crucially, the local electronic environment varies smoothly with interlayer registry. This can be seen in at least two ways. First, interlayer electron hoppings are often taken to be continuous functions of local geometry in tight binding models which provide good results in the Moir\'e approximation \cite{bistritzer_moire_2011, carr_electronic-structure_2020, carr_exact_2019, fang_ab-initio_2015, koshino_effective_2020}. More directly, Ghadimi et al. have observed the local density of states to vary more smoothly across interlayer registry than over physical space \cite{ghadimi_quasiperiodic_2025}. This smoothness implies that the local electronic environment need not be evaluated independently at every possible registry. Instead, it can be interpolated from limited set of local configurations (orange points in Fig.~\ref{fig:MatEx}(b)).

At the same time, the electronic structure is \emph{nearsighted} \cite{kohn_density_1996, ismail-beigi_locality_1999, prodan_nearsightedness_2005}: it depends primarily on the local atomic configuration. The principle of nearsightedness allows us to represent each local environment within the MINT framework \cite{niedzielski_unmasking_2025,gerber_ab_2020} as the convergent limit of sequence of finite proxy systems that progressively embed the local configuration within larger structural contexts. Fig.~\ref{fig:MatEx}(c) shows three members of the proxy sequence that converge towards bulk behavior at the given interlayer registry. Importantly, because convergence over a proxy sequence is predictable, only a subsequence of structures is needed to extrapolate the local electronic environment to its bulk limit. Thus, we require only a few small proxy systems at relatively few interlayer registries to represent the electronic structure of an incommensurate bilayer.

To do this quantitatively, we employ the use of Wannier functions. Wannier functions are constructed by Fourier transforming a subset of \emph{ab inito} electronic eigenstates. Through appropriate choice of gauge, they can be made maximally localized \cite{souza_maximally_2001}. It is actually this localization from which the nearsightedness of the electron density follows \cite{kohn_density_1996, ismail-beigi_locality_1999, prodan_nearsightedness_2005}. Finally, perturbations to Wannier functions induced by distant structural or potential changes decay with separation \cite{lihm_wannier_2021}. 

Importantly, the exact eigenstates and energies of the \emph{ab inito} Hamiltonian are reproduced in a chosen energy window by diagonalizing the Hamiltonian projected onto the Wannier basis. Moreover, this Hamiltonian inherits the same localization and insensitivity properties from the Wannier functions, allowing us to recover the full incommensurate Hamiltonian from small proxy system calculations at a limited number of registries.

\begin{figure}
\includegraphics[width=\linewidth]{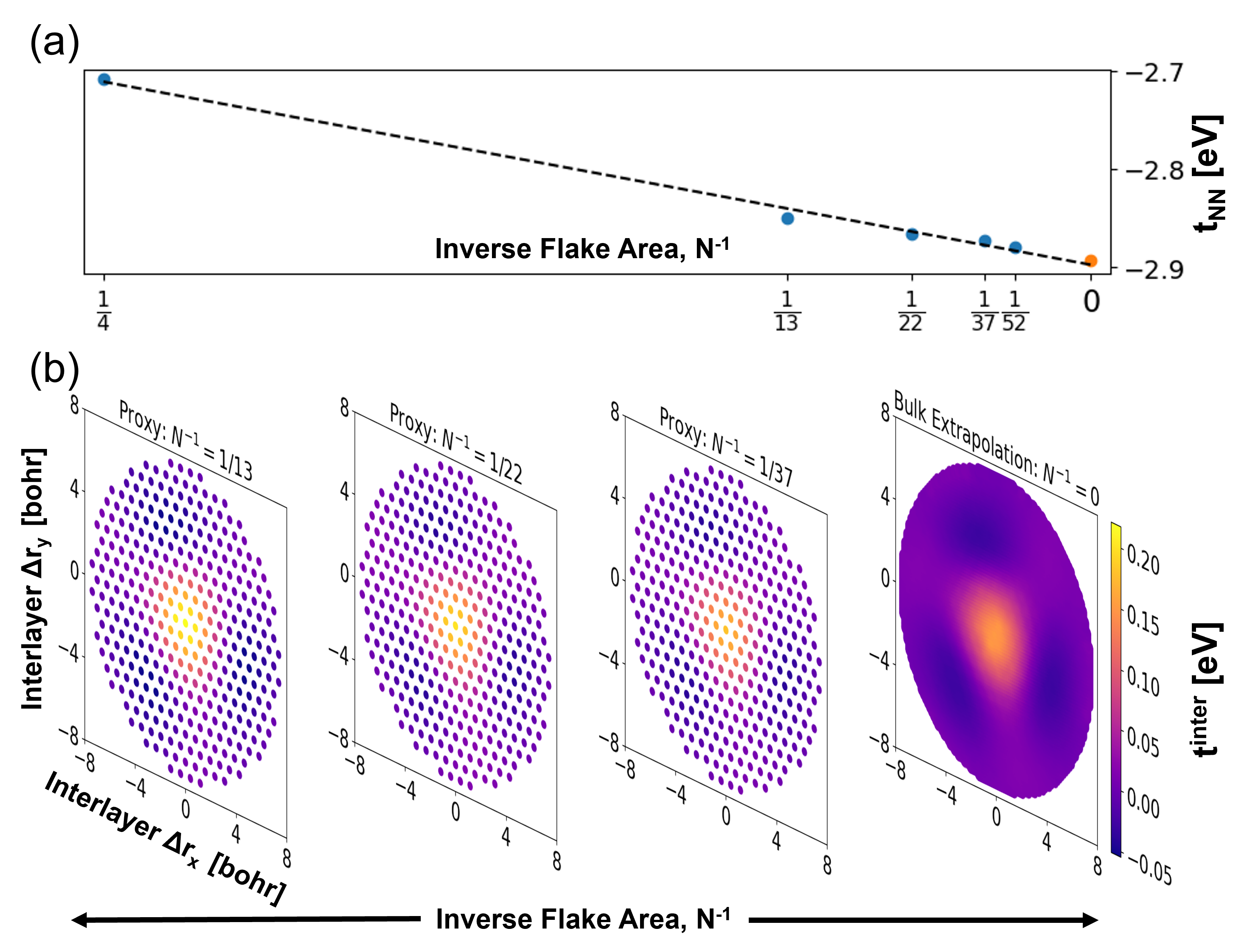}
\caption{Power-law extrapolation of Wannier matrix elements to the bulk limit (a) First nearest neighbor Wannier matrix elements for monolayer graphene proxy systems versus the inverse number of carbon atoms $N^{-1}$. Matrix elements obtained from graphene proxy systems (blue dots) show a clear power-law convergence trend with system size (dashed line fit) to the bulk graphene matrix element of $-2.89$ eV (orange dot), demonstrating the validity of this approach. Color indicates magnitude of the Wannier matrix element. (b) DFT calculated interlayer Wannier matrix elements from MINT proxy systems plotted as a function of the radial displacement and inverse flake size. Bulk values shown in the right panel are the results of a simultaneous interpolation and extrapolation using anisotropic local linear regression as described in the main text.}
\label{fig:InterpExtrap}
\end{figure}

We verify the nearsightedness of the Wannier Hamiltonian matrix elements directly for a monolayer graphene proxy sequence by computing the nearest-neighbor intralayer matrix element for flakes of increasing size. As shown in Fig.~\ref{fig:InterpExtrap}(a), the extracted matrix elements converge toward the bulk graphene value of $-2.89\,\mathrm{eV}$ with inverse-system-size scaling. This scaling is consistent with the algebraic decay expected in gapless two-dimensional systems, where boundary-induced perturbations diminish with increasing flake radius. Crucially, this power-law convergence can be used to extrapolate Wannier matrix elements from a truncated proxy sequence to their bulk value analogous to our MINT framework \cite{niedzielski_unmasking_2025}.

\subsection{Registry-Dependent First-Principles Hamiltonians}

We now describe the construction of the Wannier Hamiltonian for an incommensurate bilayer system:
\begin{equation}
\begin{split}
H = & \sum_{i,j \in A} t_A^{intra}(\mathcalligra r(r_i), \Delta r_{ij}) c_j^\dagger c_i + \\ & \sum_{i,j \in B} t_B^{intra}(\mathcalligra r(r_i), \Delta r_{ij}) c_j^\dagger c_i + \\ & \sum_{i\in A, j\in B} t^{inter}(\Delta r_{ij}) c_j^\dagger c_i ,
\end{split}
\end{equation}
where $c_i^\dagger$ and $c_i$ are the creation and annihilation operators for an electron in the Wannier orbital centered at position $r_i$, $t^{\mathrm{intra}}_{A,B}$ and $t^{\mathrm{inter}}$ are interlayer registry dependent matrix elements for intra- and interlayer hoppings, and $\Delta r_{ij}$ is the displacement between two Wannier orbitals centers. To determine $t^{\mathrm{intra}}_{A,B}$ and $t^{\mathrm{inter}}$ we combine the above notions of smooth local environments and predictable convergence by performing simultaneous interpolation and extrapolation of the proxy Wannier functions. 

We first obtain a geometric representation of the interface by constructing finite-flake substrate MINT proxy systems (Fig.~\ref{fig:MatEx}(c)) for a sufficiently dense sampling of interlayer registries, orange dots in Fig.~\ref{fig:MatEx}(b). The tedious step of generating these initial unrelaxed structures is automated by our \textsc{MINpuT} code available as an auxiliary script in Quantum ESPRESSO \cite{giannozzi_quantum_2009, giannozzi_advanced_2017}. Each of these proxy systems is then relaxed \emph{ab initio} and Wannier functions are extracted over the energy window of interest (see Methods). Importantly, to establish a consistent gauge across the system, we pin the Wannier orbitals at atomic or bond centers. 

From each relaxed proxy, we extract only the Wannier Hamiltonian matrix elements associated with orbitals near the flake center (see Methods), where boundary-induced perturbations are the smallest. Along with these matrix elements, we record the local interlayer registry and the relative displacements between orbital pairs. Repeating this procedure for all proxy systems yields intra- and interlayer matrix elements
\begin{equation}
t_{ij}^\mathrm{intra}(N^{-1}, \mathcalligra r(r_i),\Delta r_{ij}) \text{ and } t_{ij}^{\mathrm{inter}}(N^{-1}, \Delta r_{ij}),
\end{equation}
where $N$ denotes the number of atoms in the proxy system's finite flake, $\mathcalligra r(r_i)$ the local registry, and $\Delta r_{ij}$ the displacement between the two Wannier functions. Note for $t^\mathrm{inter}$ that the interlayer registry is implicitly encoded in $\Delta r_{ij}$ and for $t^\mathrm{intra}$ that $\Delta r_{ij}$ is clustered into sets of $n^{th}$ nearest neighbors which we treat independently.

Rather than interpolating over registry and extrapolating over system size as separate steps, we perform both simultaneously through local regression in this higher-dimensional space (Fig.~\ref{fig:InterpExtrap}(b)). Formally, we obtain the matrix elements in the bulk limit by evaluating this higher-dimensional interpolation at $N^{-1}=0$:
\begin{equation}
t^{\infty}(\mathcalligra r(r_i), \Delta r_{ij})) =\mathcal{R}[\{t_{kl}^{proxy}\}](N^{-1}=0, \mathcalligra r(r_i), \Delta r_{ij})),
\end{equation}
where $\mathcal{R}$ is the local regression and ${t^{\mathrm{proxy}}_{kl}}$ denotes the set of all relevant proxy derived matrix elements (see Methods). In this work we use local linear regression with a kernel that has gaussian shape along the two spatial dimensions and a flat constant value along the inverse system size dimension (effectively performing a linear fit over inverse system size). 

Finally, to establish a direct bridge between our incommensurate Hamiltonians and experimentally accessible quantities, such as angle-resolved photoemission spectroscopy (ARPES), we evaluate the momentum-resolved density of states,
\begin{equation}
n(\mathbf{q},E) = \sum_i \big| \psi_i(\mathbf{q}) \big|^2 \, \delta(E-\epsilon_i),
\end{equation}
which remains well defined even in the absence of discrete translational symmetry and thus provides a natural observable for incommensurate systems. In the present work, we compute this quantity for the 1672-atom approximant in Fig.~\ref{fig:MatEx}(a) using the reconstructed Wannier Hamiltonian (details in Methods), thereby avoiding explicit \emph{ab initio} calculations at the scale of the full system. This particular implementation demonstrates the feasibility of our approach while leaving open the use of alternative electronic-structure approaches in the future.

\subsection{Reproducing the Electronic Structure of 30$^\circ$-TBG}

\begin{figure*}
\includegraphics[width=0.9\linewidth]{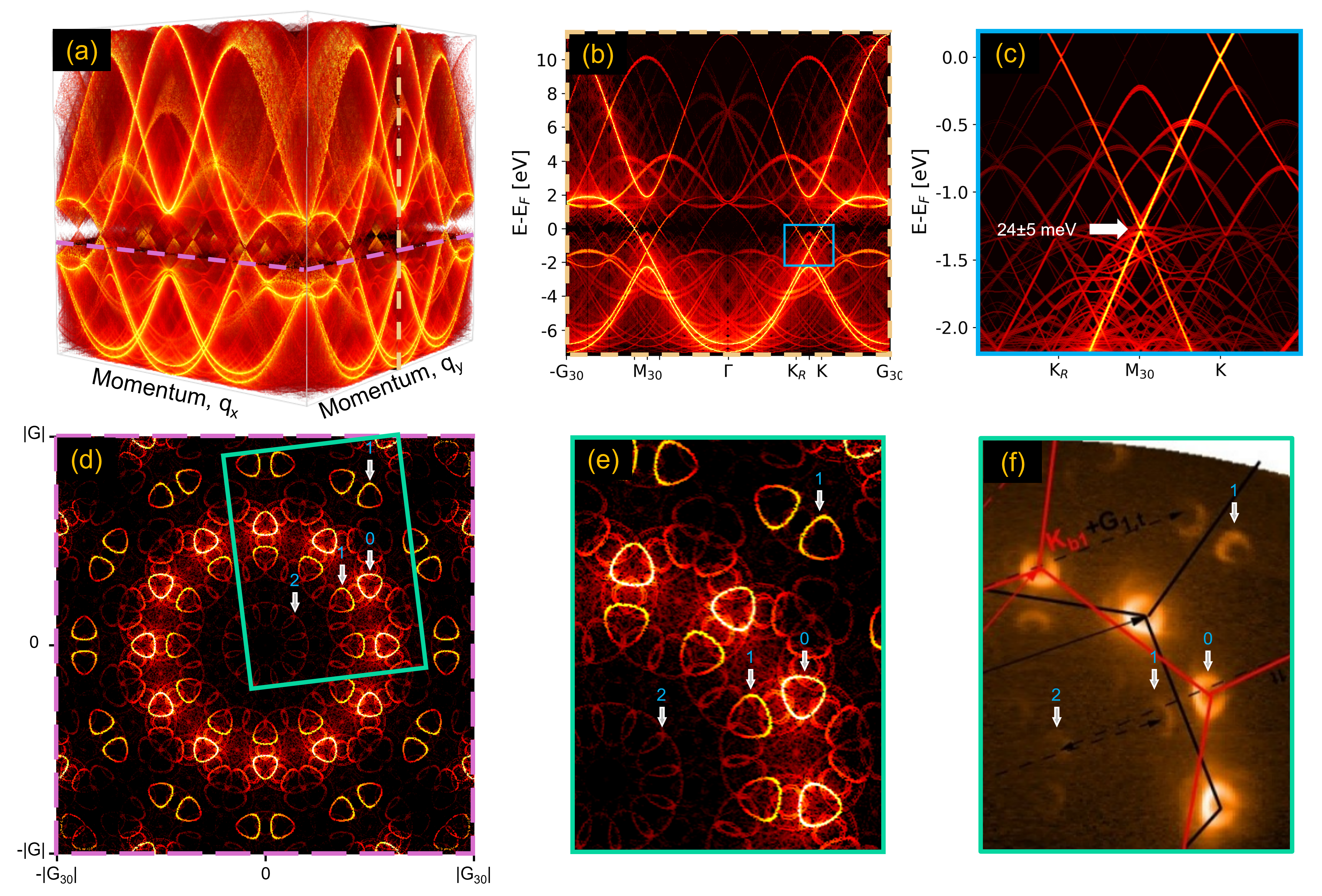}
\caption{Twelve-fold symmetric electronic structure and mirrored Dirac cones of $30^\circ$ twisted bilayer graphene. Electronic structure of $30^{\circ}$-TBG. (a) 3D Rendering of the momentum resolved DOS of 30$^\circ$-TBG calculated as described in the main text. Intensity is indicated by color hue and is plotted on a log scale to help visualize subtle features. Partial transparency is used for visualization purposes. (b) Slice through (a) at $q_y = 0$ showing the bandstructure. In addition to the regular Dirac cones at K and K', mirrored Dirac cones appear at K$_R$ and K'$_R$ as expected. (c) Zoom-in of blue-box region of (b) with higher sample density. Energy and momentum windows chosen to match the experimental Fig.~3D from Yao et al. \cite{yao_quasicrystalline_2018}. Shows expected intensity profile of main and mirrored Dirac cones as well as a minigap of $24 \pm 5$ meV where the main and mirrored cones cross at M. (d) Slice through (a) at $E-E_F=-0.8~eV$ shows the expected 12-fold rotational symmetry of 30$^\circ$-TBG shows not only main and mirrored cones, but also higher order (weaker) features such as near $\Gamma$.  (e) Zoom-in of (d) to match (f) reprint of Fig S2 (lower right panel) from Hamer et al. (2022) under a 
CC-BY 4.0 license. Displays log of ARPES intensity at $\sim0.8$~eV below the Fermi level. Original image is cropped and accent arrows have been added.}
\label{fig:MirroedDiracCones}
\end{figure*}

We now apply the locally constructed Wannier Hamiltonian framework to 30$^\circ$-TBG. Fig.~\ref{fig:MirroedDiracCones}(a) presents the logarithm of the full three-dimensional momentum-resolved density of states of 30$^\circ$-TBG, rendered to visualize the rich spectral structure arising from its quasicrystalline interlayer coupling. The calculation displays both the primary band features inherited from the individual graphene layers and numerous weaker replica states generated by quasiperiodic scattering processes. The spectrum shown was constructed from approximately $3.9\times10^{9}$ histogram samples (see Methods). Notably, once the Wannier Hamiltonian was constructed from first-principles proxy calculations, the full evaluation of $n(\mathbf{q},E)$ required only $\sim17$ minutes of GPU time, highlighting the computational efficiency of the present framework.

Fig.~\ref{fig:MirroedDiracCones}(b) shows a representative slice through $n(\mathbf{q},E)$ at $q_y=0$, displayed in the conventional energy--momentum format. In addition to the primary Dirac cones inherited from the individual graphene layers at the $K$ and $K'$ points, the calculation clearly reproduces the mirrored Dirac cones at the expected $K_R$ and $K'_R$ locations. A magnified view of this region, shown in Fig.~\ref{fig:MirroedDiracCones}(c), is computed using denser momentum sampling and plotted over an energy--momentum window chosen to coincide with ARPES measurements of 30$^\circ$-TBG on a Pt substrate reported by Yao \emph{et al.}~\cite{yao_quasicrystalline_2018}. The calculation captures both the relative positioning and intensity asymmetries of the primary and mirrored cones. Moreover, we resolve a minigap of $24\pm5$ meV at the intersection between the main and mirrored cones near $-1.4$ eV, in good visual agreement with the gap reported experimentally in Figure 5b.4 from Hamer et al.~\cite{hamer_moire_2022}.

Further comparison with experiment is obtained from constant-energy spectral maps. Figs.~\ref{fig:MirroedDiracCones}(d,e) show our momentum-resolved density of states at a constant energy of $0.8$ eV below $E_F$. This corresponds to the reported binding energy of the logarithmic ARPES map reported in Fig.~\ref{fig:MirroedDiracCones}(f) by Hamer \emph{et al.} (2022)~\cite{hamer_moire_2022}, here reprinted under a CC-BY 4.0 license. Our calculation reproduces the characteristic twelvefold rotational symmetry of 30$^\circ$-TBG, including the horseshoe-shaped primary Dirac features (labeled “0”) and their mirrored first-order replica cones (labeled “1”). These first-order features appear at the correct locations with reduced spectral intensity and crescent-shaped profiles consistent with experiment. Additionally, the calculation resolves second-order scattering features near the $\Gamma$ point (labeled “2”) arising from higher-order interlayer Umklapp processes, which naturally emerge from our construction of the Hamiltonian. While fewer such features are visible in the experimental data, this difference is consistent with the surface sensitivity of ARPES measurements, which preferentially probe the upper graphene layer. The close agreement of both primary and higher-order spectral features with experiment demonstrates the ability of the present framework to reproduce the subtle features of a quasiperiodic electronic structure.

\begin{figure}
\includegraphics[width=\linewidth]{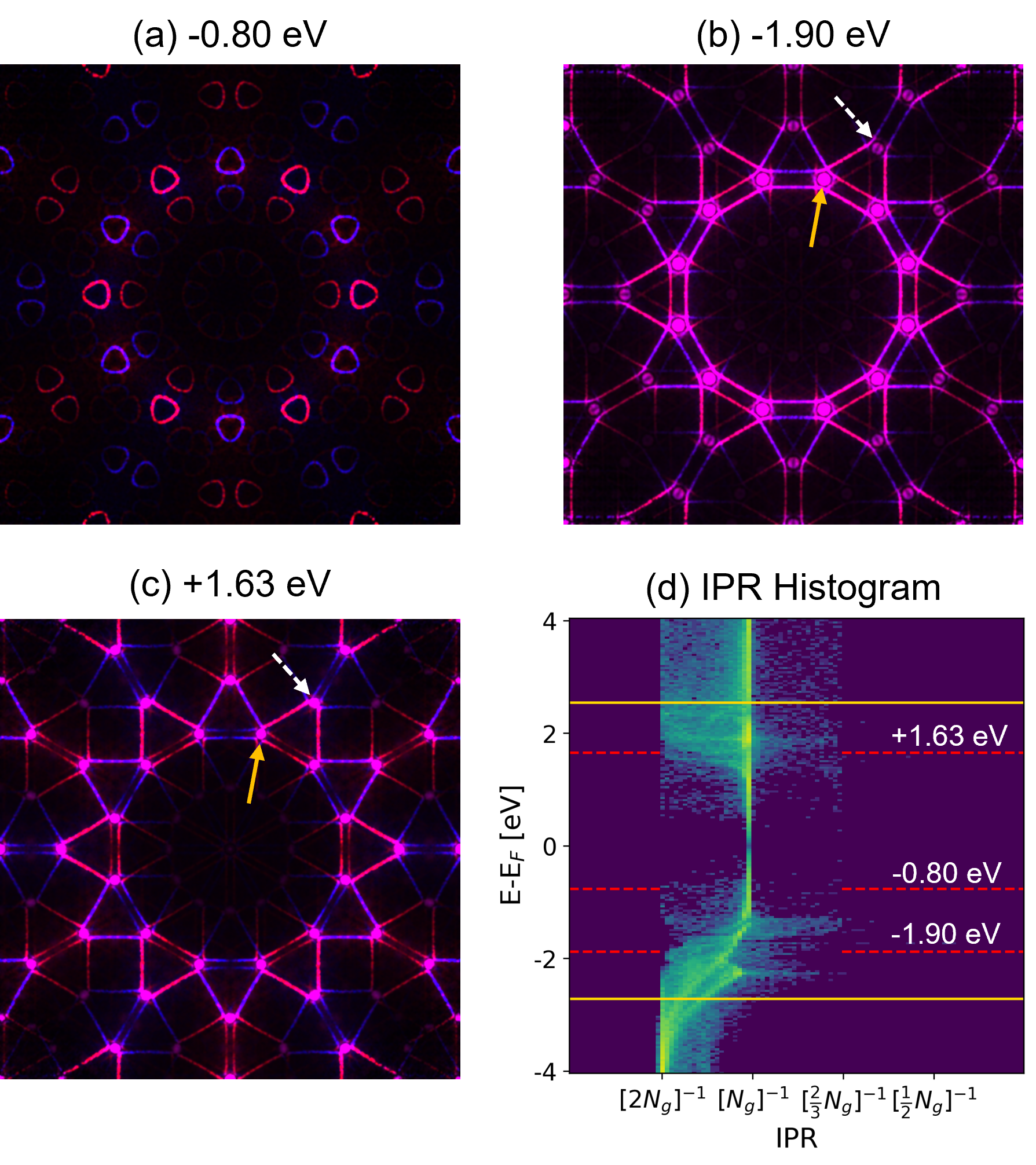}
\caption{Interlayer hybridization and localization of quasiperiodic flat-band states in $30^\circ$ twisted bilayer graphene. (a-c) Layer projected momentum resolved DOS with red(blue) hue corresponding top(bottom) layer occupation. RGB coloring is used such that states hybridized equally across both layers appear magenta. Color intensity corresponds to the square-root of the total spectral intensity to preserve relative layer-weight contrast. (a) slice at -0.80 eV shows Dirac cones and their mirrors originate primarily from a single layer. (b,c) slices at -1.90 eV and 1.63 eV displaying the formation of circular flat band features (magenta circles at solid-yellow and dashed-white arrows), at the intersection of Dirac cones, with high densities of states and strong interlayer hybridization. In (b) inner ring of flat bands (solid-yellow arrow) appears more intense than scattered replicas (white-dashed arrow), while both have similar intensities in (c). (d) Histogram of eigenstates binned by their energy and inverse participation ration (IPR). Color corresponds to the logarithm of the bin counts to accentuate subtle minority features. Yellow lines show the extent of the Wannier inner window.}
\label{fig:ProjIPR}
\end{figure}

While the momentum-resolved density of states establishes quantitative agreement with ARPES measurements, spectral intensity alone does not fully characterize the nature of the underlying electronic states. In particular, it is essential to distinguish genuinely quasiperiodic states—arising from strong hybridization between the two graphene layers—from states that remain effectively periodic and largely confined to a single layer. This distinction is central to understanding flat-band formation and its possible connection to enhanced superconductivity in quasicrystalline systems~\cite{ghadimi_quasiperiodic_2025}.

\subsection{Quasicrystalline Hybridizaiton and Localization}

To distinguish these classes of states, we analyze both the layer-resolved character and real-space localization of the electronic states of 30$^\circ$-TBG. Quasiperiodic states are expected to exhibit substantial hybridization between the two graphene layers, whereas effectively periodic states remain predominantly confined to a single layer \cite{ghadimi_quasiperiodic_2025}. We therefore compute layer-projected spectral weights by projecting each eigenstate onto Wannier orbitals associated with the top and bottom graphene layers, thereby decomposing $n(\mathbf{q},E)$ into layer-resolved components (see Methods). In parallel, we evaluate the inverse participation ratio (IPR)—a standard measure of real-space localization that quantifies the effective number of sites participating in a given state—to characterize its spatial extent and degree of bilayer mixing (see Methods).

Fig.~\ref{fig:ProjIPR}(a) shows the constant-energy slice at $-0.8$ eV colored according to layer-resolved character, with red and blue indicating states predominantly localized on the top and bottom graphene layers, respectively, and magenta indicating significant bilayer hybridization. At this energy, both the primary Dirac cones and their mirrored first-order replicas exhibit predominantly single-layer character, consistent with prior experimental observations \cite{deng_interlayer_2020}. This conclusion is reinforced by the inverse participation ratio data shown in Fig.~\ref{fig:ProjIPR}(d), where the majority of states at this energy occupy approximately one graphene layer’s worth of sites, $N_g$. We observe this predominantly single-layer character within an energy window of approximately $\pm 1$ eV around the Fermi level. Despite the absence of global translational symmetry, the Dirac states in this regime are localized to single layers and thus remain effectively periodic in character.

At energies further from the Fermi level, this single-layer character breaks down. Figs.~\ref{fig:ProjIPR}(b,c) show constant-energy contours at energies where Dirac cones originating from opposite graphene layers intersect and hybridize (solid-yellow and dashed-white arrows). At these intersection points, bright circular features emerge, corresponding to flat-band states with substantial spectral weight on both layers, indicated by their magenta coloration. In addition to the twelve dominant flat-band disks, secondary replica rings appear at larger momenta, suggesting higher-order interlayer Umklapp scattering processes analogous to those responsible for the mirrored Dirac cones. The inverse participation ratio data in Fig.~\ref{fig:ProjIPR}(d) simultaneously reveal a broad distribution of states with participation ratios approaching $2N_g$, exceeding what is possible for states confined to a single layer. The concurrent appearance of enhanced IPR and strong bilayer spectral weight confirms that these flat bands represent genuinely quasiperiodic states arising from interlayer hybridization rather than weakly perturbed monolayer states. Consistent with prior reports \cite{hamer_moire_2022, ghadimi_quasiperiodic_2025}, such states become prominent more than $\sim1$ eV below $E_F$. Notably, we also predict analogous quasiperiodic flat-band states above the Fermi level, which may be experimentally accessible through cation intercalation~\cite{mcchesney_extended_2010, kawaguchi_possible_2023, grubisic-cabo_quasi-freestanding_2024}.

Taken together, these results demonstrate that the present first-principles Wannier-Hamiltonian framework reproduces experimentally observed spectral features of 30$^\circ$-TBG, including mirrored Dirac cones and hybridization-induced minigaps, while resolving the layer character and spatial extent of the underlying electronic states. The ability to distinguish effectively periodic Dirac states from genuinely quasiperiodic flat-band states confirms that our registry-resolved Hamiltonian construction captures both the momentum and real-space structure of the incommensurate interface with fidelity.

\section{Discussion}

We have described a transferable first-principles framework for constructing the Hamiltonian of arbitrary incommensurate bilayer interfaces based on the smooth registry dependence and the nearsightedness of Wannier functions. We have shown how a limited set of \emph{ab initio} calculations on small commensurate proxy systems suffices to interpolate and extrapolate the matrix elements required to assemble the fully incommensurate Hamiltonian. We then validated our framework by reproducing the subtle, experimentally observed electronic structure of $30^{\circ}$-TBG, including mirrored Dirac cones, their higher-order replicas, and minigaps. Finally, we calculated quantities important to quasicrystalline superconductivity including interlayer hybridization and IPR. Our framework not only achieves crystalline-level accuracy in a genuinely quasiperiodic system but further establishes a foundation for extensions beyond the present case.

Here we employed one particular large commensurate supercell to determine the global geometry used in constructing the Hamiltonian, but the interpolation-based framework itself is independent of any specific approximant. Once the registry-resolved matrix elements are obtained, the resulting first-principles Hamiltonian can be assembled into arbitrarily large structures with no additional DFT cost. In principle, it can be treated directly in the thermodynamic limit using other methods that also exploit electronic nearsightedness. Techniques based on moment expansions or statistical Green’s-function approaches over relative registries could provide alternative systematically improvable routes for describing incommensurate systems, without requiring the brute-force diagonalization of large commensurate supercells.

Here, we have demonstrated our framework on $30^{\circ}$-TBG, a well-known quasicrystalline benchmark, however, our framework is not restricted to graphene nor even to bilayer systems. Multilayer stacks could be treated by resolving each interface individually or, more generally, by extending the dimensionality of registry space to incorporate the combined interlayer registries of multiple adjacent layers and interpolating over this higher-dimensional manifold. The essential requirement is the existence of a smooth local-environment descriptor, analogous to interlayer registry, over which the Hamiltonian matrix elements vary smoothly. Whenever such a description can be defined, the same interpolation-extrapolation strategy may be applied. This perspective opens a pathway toward truly first-principles treatments of lattice-mismatched multilayers, multiple twist quasicrystalline stacks, and, in principle, even amorphous solids. 

\section{Methods}

\subsection{Local Interlayer registry}

To account for local strain in an incommensurate system relative to its unstrained construction, we define the local interlayer registry for an atom at position $r_A$ in layer A by averaging over its nearest neighbors in the opposite layer, B. For each nearest neighbor at position $r_B$, we determine its \textit{local origin} $o(r_B)$ by subtracting the displacement that maps the origin of the unstrained lattice of layer B to the atomic basis position corresponding to $r_B$. We then compute the in-plane displacement from this local origin $o(r_B)$ to the atom of interest at $r_A$. This provides a good estimate of the local interlayer registry, $\mathcalligra{r}(r_A)$, associated with the neighbor at $r_B$. Finally, we average these in-plane displacements over all nearest neighbors in layer B to obtain the strain-robust local interlayer registry at $r_A$.

\subsection{DFT Relaxation}

All \emph{ab initio} calculations in this work were performed using the JDFTx software package \cite{sundararaman_jdftx_2017}, using ultrasoft GBRV pseudopotentials \cite{garrity_pseudopotentials_2014} and the generalized gradient approximation (GGA) to the exchange-correlation functional \cite{perdew_generalized_1996}. Van der Waals interactions were treated with Grimme D3 corrections to improve the accuracy of interlayer forces~\cite{grimme_consistent_2010}.

Brillouin zone sampling for the MINT proxy calculations was chosen to maintain a reciprocal-space $k$-point density at least as high as that of monolayer graphene sampled with a $\Gamma$-centered $24 \times 24 \times 1$ mesh. This corresponds to $6 \times 6 \times 1$, $5 \times 5 \times 1$, and $4 \times 4 \times 1$ meshes for the C$_{13}$H$_9$, C$_{22}$H$_{12}$, and C$_{37}$H$_{15}$ system sizes, respectively. We employ a Fermi--Dirac smearing of $0.001$ Hartree for the electron occupations to better reflect experimental conditions and facilitate convergence. A vacuum spacing of 25 Bohr radii was used in the out-of-plane direction, together with Coulomb truncation, to minimize spurious interactions between periodic images.

Electronic minimizations were performed to within a tolerance of 0.01 mev/atom. Lattice and ionic coordinates were relaxed to a tolerance 0.1 meV/atom.

\subsection{Wannier Function Calculations}

Obtaining localized and consistently centered Wannier functions is crucial for the framework we propose in this work. Those readers intending to use our framework should therefore put great care into this step.

Wannier functions for each of our proxy systems were calculated with the Wannier package of JDFTx \cite{sundararaman_jdftx_2017}. We initialized our Wannier optimization with carbon pz orbitals centered on each carbon atom in the proxy system. Wannier functions were pinned to the carbon atomic centers by penalizing deviations during Wannier spread minimization. We employed the inner-outer window approach, using an inner window of $[-2.73, 2.51]$ eV around $E_F$ to enclose the parts of the bandstructure where only carbon pz orbitals contribute and extended a wide outer window of $[-8.5,13]$ eV to encompass the full energy range where these orbitals contribute. Wannier function localization was treated using a real-space measure of Wannier function spread.

\subsection{Central Region Wannier Functions}

To avoid effects from the boundary of the flake in a MINT proxy system, we only record Hamiltonian matrix elements for Wannier orbitals near the center of the flake. Practically, we define a source radius $R_{s}$ and a target radius $R_{t}$ from the center of the flake. We record the matrix element $t_{ij}$ (between orbitals centered at $r_i$ and $r_j$) only if the in-plane component of $r_i$ lies within the source radius and $r_j$ lies in the target radius. Our calculations use $R_{s} = 3.0$ Bohr radii and $R_{t} = 7.5$ Bohr radii.

\subsection{Local Linear Regression}

We perform local linear regression (LLR) using a kernel having gaussian shape (width 0.25 Bohr radii) over the two spatial dimensions and a constant value over the inverse system size dimension. The effect is a smoothed interpolation over the interlayer registry axes and a linear fit along the inverse system size direction. In practice, we perform separate interpolations between each pair of symmetrically unique orbitals, e.g. Wannier matrix elements between the upper graphene layer's A sublattice and the lower graphene layer's B sublattice. Intralayer hoppings are grouped into sets of nearest neighbors according to their $\Delta r_{ij}$ before interpolation/extrapolation.

\subsection{$30^\circ$-TBG Hamiltonian Construction}

Here, for $30^\circ$-TBG, we construct a Wannier Hamiltonian for the fourth commensurate approximant to $30^\circ$-TBG (Fig.~\ref{fig:MatEx}(a)) containing 1672 atoms. We obtain the first, second, and third nearest neighbor matrix elements by extrapolating the corresponding elements from periodic substrates in our MINT proxy calculations averaged over all interlayer registries (we find very little difference in the interlayer matrix elements wrt interlayer registry for $30^\circ$-TBG, but we note that for systems where these differences are larger, and better resolved by DFT, that interpolation can be performed over them as is done with interlayer elements). While our method can in principle, given enough compute time, be used to obtain nearest neighbor elements arbitrarily far away, we chose to set the 4th and 5th nearest neighbor elements to those of bulk graphene as these distances are large compared to the flakes we used in our proxy systems.  We obtain the interlayer matrix elements for interlayer displacements with a plane-project radius less than 4 Bohr radii, as their magnitudes are less than 25meV beyond this cutoff.

We sampled 100 kpoints in the Brillouin zone of the commensurate approximant system (CBZ). A Hamiltonian diagonalization for a given kpoint yields 1672 electronic eigenvalues and eigenstates. Each eigenvalue is unfolded to a 43x45 grid of RLVs of the large commensurate approximant system. For Figs.~\ref{fig:MirroedDiracCones}(a,b,d,e) in the main-text, we further use the 12-fold rotational symmetry to obtain 12 times this sampling via rotational averaging. This results in $2,322,000$ momenta and $3.88\times10^9$ total histogram samples for Figs. 3(a,b,d,e).

Hamiltonian diagonalizations and other postprocessing was done using the CuPy library \cite{cupy_learningsys2017} for GPU accelerated Python. Computations were performed on an NVIDIA RTX 4500 Ada Generation GPU with the 3D histogram in Fig.~\ref{fig:MirroedDiracCones}(a).

\subsection{Layer Projections}

The layer participation fraction $f_A$, for a state $\psi$ projected onto layer A, is
\begin{equation}
f_A(\psi) = \sum_{i \in A} |\braket{w_i | \psi}|^2,
\end{equation}
where the sum is performed over all Wannier orbitals $w_i$ belonging to layer A. The layer-resolved momentum resolved DOS is then
\begin{equation}
n_A(q,E) = \sum_i f_A(\psi_i) |\psi_i(\mathbf{q})|^2 \delta(E-\epsilon_i).
\end{equation}

We calculated $n_\mathrm{top}(q,E)$ and $n_\mathrm{bot}(q,E)$ for the top and bottom layers employing the same calculation parameters used to obtain $n(q,E)$. Instead of averaging over a 12-fold rotational symmetry, we instead separately averaged each over a 6-fold rotational symmetry. 

Figs.~\ref{fig:ProjIPR}(a-c) plot the square-root of the total spectral intensity while keeping the layer mixing fractions consistent for each histogram bin by using $\frac{n_{top}(q,E)}{n(q,E)} n(q,E)^{1/2}$ for the red channel and $\frac{n_{bot}(q,E)}{n(q,E)} n(q,E)^{1/2}$ for the blue channel. Note that such a construction is not possible for logarithms of the total intensity.

\subsection{Inverse Participation Ratio}

The inverse participation of a normalized eigenstate $\psi$ in the Wannier basis reduces to a sum over the Wannier orbtials, $w_i$:\
\begin{equation}
\mathrm{IPR}(\psi) = \sum_i^{N} |\braket{w_i |\psi}|^4 ,
\end{equation}
where N is the total number of Wannier orbitals modeled. In this definition, a perfectly de-localized state has $\mathrm{IPR} = 1/N$ and a state localized entirely on a single Wannier orbital would have $\mathrm{IPR} = 1$.

The inverse participation ratio histogram shown in Fig.~\ref{fig:ProjIPR}(d) is calculated monte carlo sampling the Hamiltonian with the sample density as was used for $n(q,E)$. IPR was calculated for each sampled eigenstate and along with the corresponding eigenvalue was used to create histogram samples. The histogram employs 500 bins for the energy range $[-7.5,12]$ eV (visually cropped to the range $[-4,4]$ eV) and 100 bins for the IPR range $[0,5N_g/2]$. Logarithm of histogram bin counts is shown in Fig.~\ref{fig:ProjIPR}(d).

\section{Data availability}

Data will be made available in a public repository upon publication.

\section{Code availability}

Code will be made available in a public repository upon publication.

\section{Contributions}

D.N. and T.A.A. jointly conceived the registry-space interpolation approach. D.N. developed the theoretical framework, implemented the software, and carried out the calculations. T.A.A. supervised the project. D.N. and T.A.A. jointly wrote the manuscript.

\section{Ethics declarations}

The authors declare no competing interests.

\begin{acknowledgments}
This work made use of the theory facility of the Platform for the Accelerated Realization, Analysis, and Discovery of Interface Materials (PARADIM), which is supported by the National Science Foundation under Cooperative Agreement No. DMR-2039380. D.N. acknowledges support under this agreement and T.A.A. acknowledges partial support. D.N. thanks Shake Karapetyan for her expertise and assistance with data visualizations. D.N.  acknowledges Prof. Youngkuk Kim for highlighting the $30^\circ$ twisted bilayer graphene system as a good benchmark.
 
\end{acknowledgments}

\bibliographystyle{apsrev4-1}
\bibliography{bibliography}

\end{document}